\documentclass[sigconf]{acmart}

\AtBeginDocument{%
  }

\setcopyright{acmlicensed}
\copyrightyear{2026}
\acmYear{2026}
\acmDOI{XXXXXXX.XXXXXXX}

\acmISBN{978-1-4503-XXXX-X/2026/07}

\usepackage{booktabs}
\usepackage{multirow}
\usepackage{graphicx}
\usepackage{amsmath}
\usepackage{algorithm}
\usepackage{algorithmic}
\usepackage{tikz}
\usepackage{xcolor}
\usepackage{graphicx} 
\usetikzlibrary{shapes,arrows,positioning,fit,backgrounds,shadows,calc}
\usepackage{listings}
\usepackage{tabularx}

\definecolor{llmcolor}{RGB}{144,169,223}
\definecolor{recsyscolor}{RGB}{255,183,121}
\definecolor{feedbackcolor}{RGB}{154,223,144}
\definecolor{semanticcolor}{RGB}{255,255,200}

\begin{document}

\title{RGAlign-Rec: Ranking-Guided Alignment for Latent Query Reasoning in Recommendation Systems}



\author{Junhua Liu}
\authornote{Co-first authors who contributed equally to this work.}
\email{j@forth.ai}
\affiliation{
  \institution{Forth AI}
  \country{Singapore}
}

\author{Jihao Yang}
\authornotemark[1]
\email{jihao.yang@shopee.com}
\affiliation{
  \institution{Shopee}
  \country{Singapore}
}

\author{Cheng Chang}
\email{cheng.chang@shopee.com}
\affiliation{
  \institution{Shopee}
  \country{Singapore}
}

\author{Kunrong Li}
\email{kunrong.li@forth.ai}
\affiliation{
  \institution{Forth AI}
  \country{Singapore}
}

\author{Bin Fu}
\authornote{Corresponding author.}
\email{bin.fu@shopee.com}
\affiliation{
  \institution{Shopee}
  \country{Singapore}
}

\author{Kwan Hui Lim}
\email{kwanhui@acm.org}
\affiliation{
  \institution{Singapore Uni.\ of Tech.\ and Design}
  \country{Singapore}
}


\begin{abstract}

Proactive intent prediction is a critical capability in modern e-commerce chatbots, enabling "zero-query" recommendations by anticipating user needs from behavioral and contextual signals. However, existing industrial systems face two fundamental challenges: (1) the semantic gap between discrete user features and the semantic intents within the chatbot’s Knowledge Base, and (2) the objective misalignment between general-purpose LLM outputs and task-specific ranking utilities. To address these issues, we propose RGAlign-Rec, a closed-loop alignment framework that integrates an LLM-based semantic reasoner with a Query-Enhanced (QE) ranking model. We also introduce Ranking-Guided Alignment (RGA), a multi-stage training paradigm that utilizes downstream ranking signals as feedback to refine the LLM's latent reasoning. Extensive experiments on a large-scale industrial dataset from Shopee demonstrate that RGAlign-Rec achieves a 0.12\% gain in GAUC, leading to a significant 3.52\% relative reduction in error rate, and a 0.56\% improvement in Recall@3. Online A/B testing further validates the cumulative effectiveness of our framework: the Query-Enhanced model (QE-Rec) initially yields a 0.98\% improvement in CTR, while the subsequent Ranking-Guided Alignment stage contributes an additional 0.13\% gain. These results indicate that ranking-aware alignment effectively synchronizes semantic reasoning with ranking objectives, significantly enhancing both prediction accuracy and service quality in real-world proactive recommendation systems.

\end{abstract}



\maketitle

\section{Introduction}
\label{sec:intro}

In the rapidly evolving e-commerce landscape, AI-powered chatbots have progressed from simple intent classification systems \cite{liu2024lara} to LLM-based dialogue platforms. In industrial settings, these chatbots act as the primary customer service interface, handling millions of interactions daily. Beyond response generation, a key opportunity for improving user experience lies in proactive intent prediction. Unlike recent work that emphasises dialogue management or agent-like response generation, this paper focuses on predicting a user’s underlying service bottleneck (e.g., delivery delay or payment failure) upon entry into the chatbot interface. By recommending relevant intents from a Knowledge Base (KB) in a Zero-Query setting, the system can reduce user friction and improve both click-through rate (CTR) and customer satisfaction (CSAT).

While proactive intent prediction shares surface similarities with traditional recommendation systems \cite{wide&deep}, it introduces distinct challenges. Sequential recommendation models \cite{DIN} aim to capture long-term user preferences from interaction sequences to optimize Hit Rate (HR@N) or Gross Merchandise Volume (GMV). In contrast, chatbot users are driven by short-term, context-dependent service issues rather than stable interests. The task is therefore to infer a user’s latent semantic intent within a KB under Zero-Query scenarios \cite{MINT_E}. This requires synthesizing heterogeneous signals, including order status, user profiles, and historical dialogue logs, to surface the top-$K$ candidate intents before users explicitly articulate their problems.

\begin{figure}[h]
    \centering
    \includegraphics[width=0.48\textwidth]{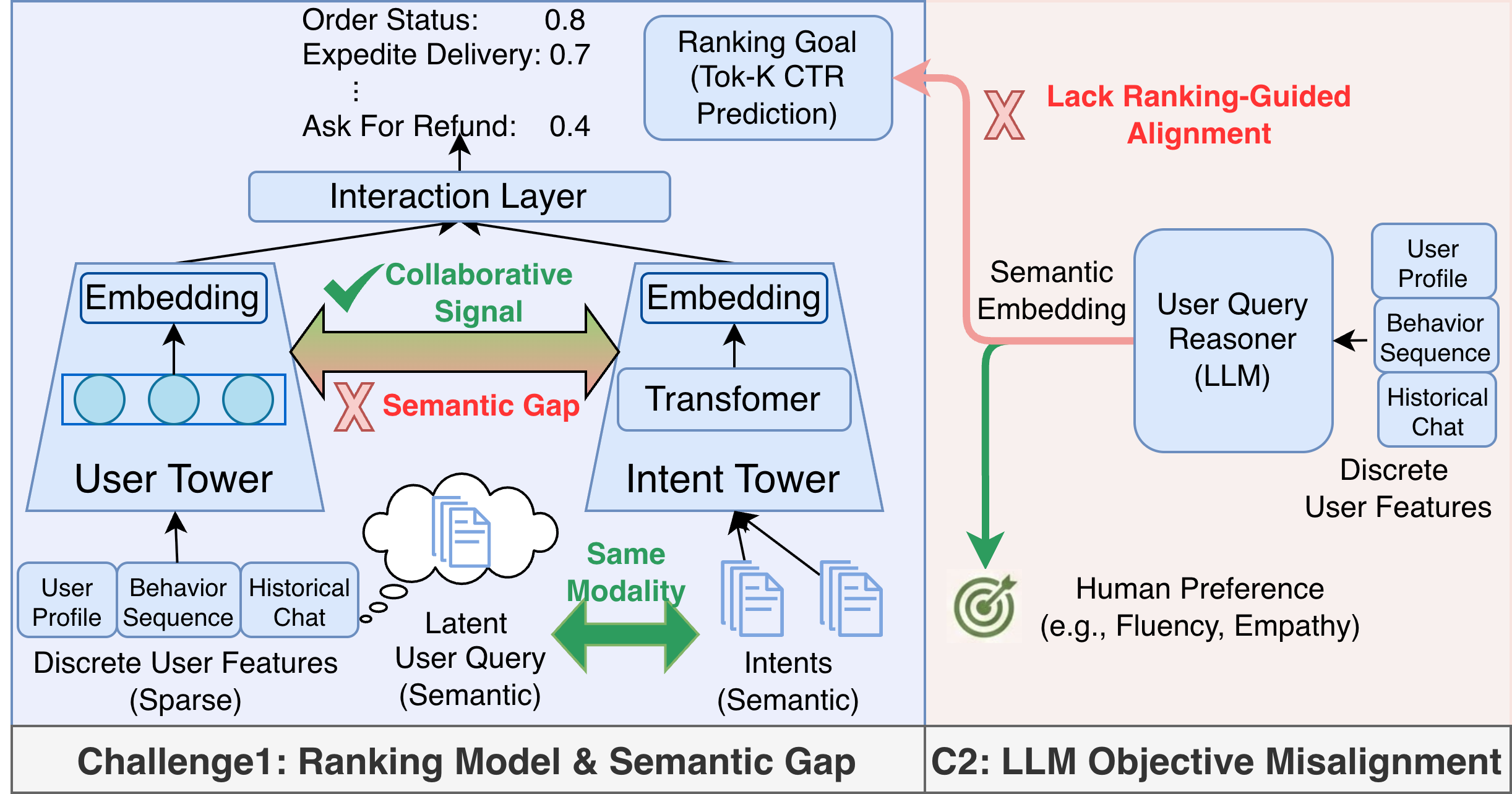} 
    \caption{Challenges of Semantic Gap and Objective Misalignment}
    \label{fig:challeges} 
\end{figure}

Despite the success of Deep Learning Recommendation Models (DLRMs) in industrial ranking systems, two fundamental challenges limit their effectiveness for proactive intent prediction, as illustrated in Figure~\ref{fig:challeges}.

\textbf{Challenge 1: Semantic Gap and Modality Mismatch.}
Existing DLRMs primarily rely on collaborative signals from discrete identifiers such as user IDs or intent IDs to optimize CTR. However, chatbot intents are text-rich and semantically fine-grained. Standard recommendation architectures often struggle to align a user’s latent service issue with textual intent descriptions in the KB, particularly in Zero-Query settings without explicit queries. Bridging the gap between sparse behavioral features and continuous semantic representations remains a central challenge.

\textbf{Challenge 2: Objective Misalignment between Human Preference and Ranking Performance.}
Recent systems increasingly incorporate Large Language Models (LLMs) as semantic feature extractors or reasoning modules \cite{hllm}. Common alignment techniques, such as supervised fine-tuning (SFT) or direct preference optimization (DPO) \cite{DPO}, are optimized for human-centric linguistic preferences, including fluency and perceived helpfulness \cite{huamnalign,llm_as_judge}. These objectives do not directly correspond to downstream ranking goals. Consequently, LLM-derived representations may be interpretable to humans but sub-optimal for improving ranking metrics such as CTR. Effective proactive services therefore require alignment mechanisms that are explicitly guided by ranking-aware feedback.

To address these challenges, we propose \textbf{RGAlign-Rec}, a framework that aligns LLM-based semantic reasoning with a query-enhanced recommendation architecture for proactive intent prediction. The framework follows a multi-stage alignment process. First, an LLM reasoner infers a latent query representation from sequential and not-sequential features, which is consumed by a Query-Enhanced ranking model (QE-Rec). Second, the trained QE-Rec model is used as a Reward Model (RM) to provide ranking-aware feedback for LLM reasoner alignment. We proposed a Ranking Guided Alignment(RGA) framework to select the best-of-$N$ query candidates generated by teacher models (GPT, Gemini, and CompassMax) for RG-supervised fine-tuning (RG-SFT) and apply in-batch contrastive learning as RG-CL over last-token-pooling embeddings to align semantic and ranking representations. Finally, the refined LLM reasoner is used to retrain the ranking model, forming a closed-loop system that consistently aligns semantic reasoning with ranking objectives.

Our main contributions are summarised as follows:
\begin{itemize}
  \item \textbf{RGAlign-Rec Architecture:} We introduce RGAlign-Rec, which integrates an LLM-based semantic query reasoner into a query-enhanced three-tower recommendation framework to bridge the semantic gap in Zero-Query intent prediction.
  \item \textbf{Closed-Loop Ranking-Guided Alignment:} We propose a Ranking-Guided Alignment(RGA) that uses the ranking model as a Reward Model for the LLM reasoner alignment, ensuring semantic representations are aligned with downstream ranking objectives.
  \item \textbf{Industrial Impact:} We validate RGAlign-Rec through offline experiments on a self-built benchmark and large-scale online A/B tests at Shopee, demonstrating consistent improvements in CTR and Customer Satisfaction (CSAT) in a production chatbot system.
\end{itemize}

\section{Related Work}
\label{sec:related}

\subsection{Generative Model for RecSys}
Recent research in recommender systems has shifted from modelling discrete feature interactions \cite{xdeepfm,dcnv2} and item sequences \cite{dien,sim} toward leveraging LLMs as feature extractors \cite{GR_category_survey} and Semantic ID (SID)-based \cite{semanticID} generative recommendation (GR) frameworks \cite{tiger}. Prior work in this area falls into two categories. \textbf{(1) LLM as feature reasoner:} models such as HLLM \cite{hllm}, NoteLLM \cite{notellm,notellmv2}, and RecGPT \cite{recgpt,recgptv2} use LLMs to distil user interests or semantic knowledge into latent representations that serve as auxiliary features for downstream recommendation models. While effective at enriching sparse behavioral signals, semantic reasoning in these approaches is typically optimized independently of ranking objectives. \textbf{(2) GR with SID:} another line of work converts item identifiers into LLM-readable semantic IDs \cite{rqvae}, reformulating recommendation as a unified generative task \cite{geng2022p5}. Methods such as \cite{tiger,hstu,han2025mtgr,deng2025onerec} employ attention and masking mechanisms to aggregate SID-based sequential information and user profiles for next-item prediction. Follow-up studies improve feature context engineering \cite{Zhang2025OneTransUF, Dai2025OnePieceBC} and attention efficiency to support industrial scalability \cite{rankmixer}. In contrast to these one-way pipelines, RGAlign-Rec introduces a ranking-guided alignment strategy that directly aligns latent user query representations with ranking objectives.

\subsection{Task-Specific Alignment via model feedback}
Large Language Models (LLMs) \cite{Achiam2023GPT4TR} are commonly aligned with human preferences using reinforcement learning from human feedback (RLHF) \cite{RLHF} or direct preference optimization (DPO) \cite{DPO}. While effective for natural language processing tasks, applying these approaches to recommendation systems is challenging due to the high cost of human annotation and the sparsity of user-item interactions. Reinforcement Learning from AI Feedback (RLAIF) \cite{Lee2023RLAIFVR, Lin2025RecR1BG} mitigates annotation cost by using automated critics, as explored in DPO-Rec \cite{DPO4Rec}, but in industrial recommendation settings, direct DPO often suffers from gradient instability and noisy supervision because discrete preference signals inadequately capture fine-grained relevance differences. Recent work such as RLMRec \cite{RLMRec} therefore shifts toward representation-level alignment, bridging collaborative signals with semantic knowledge encoded in LLMs \cite{Ye2025DASDS,Ye2025Align3GRUM}. Rather than optimizing choice probabilities, these methods align hidden representations with dense semantic embeddings. In parallel, contrastive learning (CL) \cite{CLmethod} has emerged as a more stable alternative to DPO, providing continuous optimization signals that are more robust to distribution shifts and interaction noise \cite{SDPO}. Unlike generic RLAIF-based approaches, RGAlign-Rec introduces a mutual refinement mechanism that reintegrates aligned semantic embeddings into a query-enhanced ranking model, forming a closed feedback loop that jointly supports semantic reasoning and high-performance production-scale ranking.

\section{Preliminary}
\label{sec:Preliminary}

\subsection{Problem Formulation}


Our intent prediction, like a ranking system in a cascade ranking paradigm(recall-rank), leverages heterogeneous user features and intent to recommend the top-K intents with which the user is most likely to engage. We define $z_u \in \mathbb{R}^d, z_i \in \mathbb{R}^d,$ and $z_q \in \mathbb{R}^d$ as the latent vector representations for the user, intent, and LLM-generated semantic query, respectively. The goal of our proactive intent prediction task is to learn a query-enhanced scoring function $F(z_u, z_i, z_q; \delta)$ that ranks a set of candidate intents $\mathcal{I}_c$ to maximize the probability of user interaction. Specifically, the top-K intents are selected as:

\begin{equation}
    i^* = \arg\max_{i \in \mathcal{I}_c} F(z_u, z_i, z_q; \delta)
\end{equation}
where $F(\cdot)$ denotes a query-enhanced scoring function parameterized by $\delta$. This formulation allows the model to jointly leverage collaborative user-intent signals and semantic relevance between the query embedding and candidate intents.

The system then generates a top-K ranked list  $\mathcal{I}_{r} = \text{sort}(\mathcal{I}_{c}, F_\delta)$ by sorting candidate intents in descending order based on these scores and recommending top-K intents that a user is most likely to interact with upon entering the chatbot interface. 

\subsection{LLM-Based User Query Reasoner $\pi$}
To bridge the gap between user's discrete features and semantic intents, We adopt Qwen3-4B\cite{qwen3} as User Query Reasoner to transform discrete user features into natural language descriptions, leveraging the reasoning capabilities of LLM to "synthesize" a latent query representation. The reasoning framework follows a two-stage pipeline: Feature Verbalisation and LLM semantic reasoning.

\begin{table}[htbp]
\centering
\caption{Examples of Feature Verbalization Mapping.}
\label{tab:feature_verbalization}
\scriptsize 
\setlength{\tabcolsep}{2.5pt} 
\begin{tabularx}{\columnwidth}{@{}l l p{1.8cm} X@{}}
\toprule
\textbf{Category} & \textbf{Feature} & \textbf{ID $\rightarrow$ String} & \textbf{Prompt Fragment} \\ \midrule
\textbf{Profile} & Gender & $2 \rightarrow$ Female & "User Gender: Female" \\ \midrule
\textbf{Behavior} & Intent Seq. & $[2184, \dots] \rightarrow$ [...] & "Clicked Intents In 24 Hours: ('When will I receive...', ...)" \\ \midrule
\textbf{Context} & Entry Point & $3 \rightarrow$ Shipping Page & "Entry Point: Shipping Page" \\ \cmidrule{2-4}
 & Order Stat. & $8 \rightarrow$ TO\_RECEIVE & "Current Order Status: TO\_RECEIVE" \\ \bottomrule
\end{tabularx}
\end{table}

\textbf{Feature Verbalization} 
Instead of complex architectures to handle heterogeneous features (non-sequential and sequential)\cite{huang2026hyformer,chen2025homer}, we transform feature values into a unified natural language stream. As illustrated in Table \ref{tab:feature_verbalization}, discrete feature IDs are first mapped to their corresponding categorical strings via a predefined metadata dictionary. These strings are then encapsulated into structured natural language fragments based on their logical dimensions: User Profile, Behavioral Sequence, and Interaction Context.


Unlike traditional one-hot or hash-based encoding, this verbalization preserves the inherent semantic relationships between features (e.g., the temporal urgency implied by TO\_RECEIVE for 7 days). These fragments are concatenated into a "Chain-of-Thought" (CoT) inspired user context prompt which serves as the input for the User Query Reasoner to synthesize the latent user query.
\begin{lstlisting}
##System Prompt
You are a helpful AI assistant. You first think about the reasoning process in the mind and then provide the user with the answer.

##User Prompt
You are an intelligent assistant supporting an e-commerce Chatbot. When the user enters the Chatbot via different entry points, proactively suggest one "hot question" (common intent) that the user is most likely to ask right now, based on the user related information.

Below is User Context Information, User Profile, Order Status and User Historical Behaviour information:
User Gender: Female, ... Order_Status: TO_RECEIVE

Based on the above information, generate one question that users may encounter. Please describe in one sentence the question that the user may want to ask.

Requirements: Keep the question between 10-20 words.
The user question is: <Output>
\end{lstlisting}




\textbf{LLM Reasoning Representation Learning}
We employ Qwen3-4B as semantic reasoner, taking result of feature verbalization $C$ as user context, leveraging its inherent common sense and pre-trained e-commerce knowledge to reason latent query/doubts by considering user's features into a continuous semantic representation.

A critical design choice in LLM-based recommendation systems is how to extract semantic representation from LLM outputs. Common approaches include mean pooling over all token representations~\cite{Roth2025ResourceEfficientAO}, appending an explicit end-of-sequence (EOS) token and use its hidden state~\cite{notellm} and using off-the-shelf sentence embedding models (e.g., BGE) on generated query. However, these methods are not explicitly optimized for recommendation objectives and may dilute task-relevant semantic signals. 

We adopt \textbf{Last Token Pooling (LTP)} embeddings\cite{lasttokenpooling}, which utilise the hidden state of the final token from a user context prompt to represent the user's latent query. Given feature verbalization result $C$, it is first processed by the LLM’s tokenizer to produce a token sequence $s_i = \{t_{i1}, \dots, t_{il}\}$, where $l$ denotes the sequence length. We then extract the final-layer hidden states via the LLM reasoner $\mathbf{\pi}(s_i)$, represented as 
\begin{equation}
H = [h_{i1}^{(t)}, \dots, h_{il}^{(t)}] \in \mathbb{R}^{d \times l}, 
\end{equation}
where each $h_{ij}^{(t)} \in \mathbb{R}^{d \times 1}$ is a $d$-dimensional vector. Finally, the hidden state of the last token from $H$ to derive the LTP embedding 
\begin{equation}
e_{LTP} = H[-1]. 
\end{equation}

\begin{figure*}[t]
\centering
\includegraphics[width=1\textwidth]{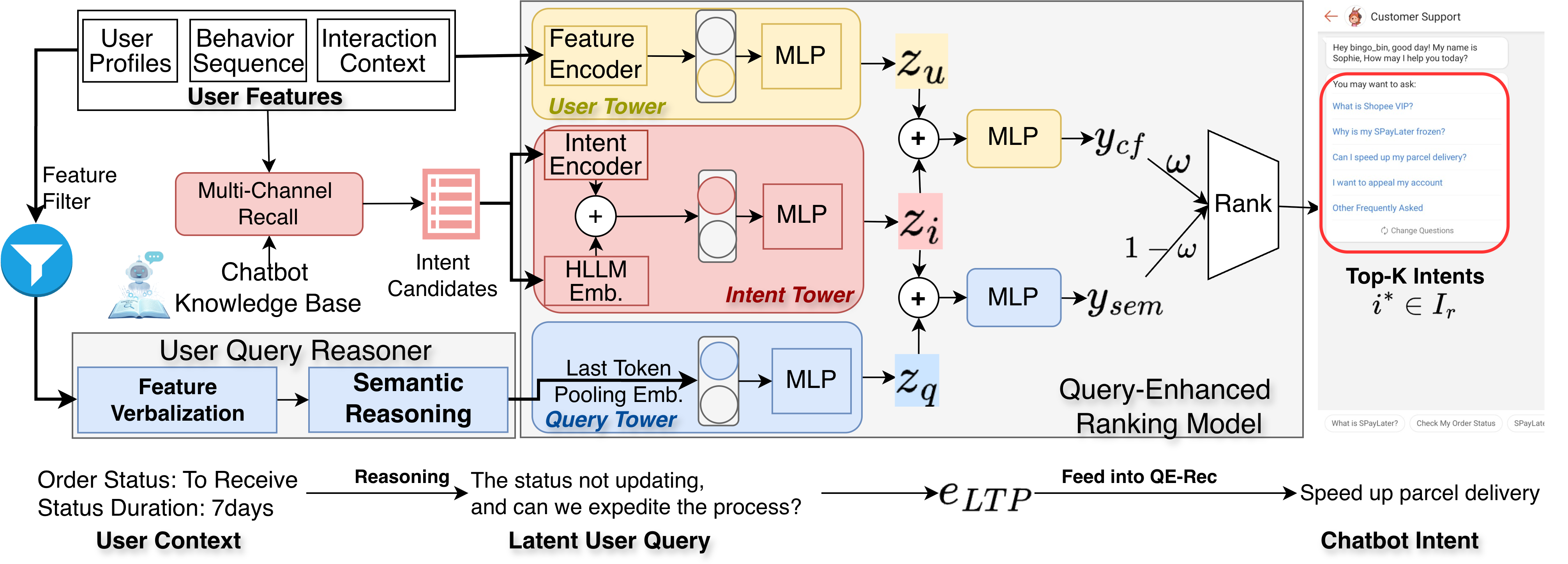}
\caption{RGAlign-Rec Framework and Online Inference Solution}
\label{fig:framework}
\end{figure*}

To illustrate the reasoning process, we present a representative workflow in the Figure \ref{fig:framework}. The LLM reasoner ingests discrete User Context like "an order remaining in the 'To Receive' status for 7 days" and reasoner it into a coherent Latent User Query: 'The status not updating, and can we expedite the process?' This transformation bridges the semantic gap by converting raw interaction logs into a natural language representation of the user’s underlying concern. This last token embedding $e_{LTP}$ then serves as a high-fidelity semantic representation, which is fed into the QE-Rec model to accurately predict the corresponding Chatbot Intent 'Speed up parcel delivery' from the Knowledge Base. This mechanism allows the system to proactively address user needs even in Zero-Query scenarios where no explicit input is provided.

\subsection{Query-Enhance Recommendation (QE-Rec)}
\label{subsec: QE-rec}


To mitigate the semantic gap within our ranking model and utilise the semantic embedding from the LLM reasoner, we extend the classical two-tower recommendation architecture~\cite{huang2013learning} with an additional query tower that explicitly models user latent queries, named Query-Enhanced Recommendation (QE-Rec).

Specifically, the proposed architecture consists of the following three towers:

\textbf{User Tower $Z_u$:} 
The User Tower employs a hierarchical encoding architecture tailored for heterogeneous features. Numerical and categorical features are encoded and concatenated into a non-sequence vector($\mathbf{v}_{non-seq}$). User historical sequences are processed via a Self-Attention mechanism in order as ($\mathbf{v}_{seq}$). For coupled feature groups (e.g., items and timestamps), we apply Intra-group Cross-Attention to model fine-grained correlations before fusing them into group-level representations$\mathbf{v}_{grouped}$. The final user representation is derived by aggregating these multi-view embeddings:$$\mathbf{z_{u}} = \sigma(\text{MLP}([\mathbf{v}_{non-seq}; \mathbf{v}_{seq}; \mathbf{v}_{grouped}]))$$This hierarchical design ensures that the model effectively differentiates between static preferences and evolving behavioral intents.

\textbf{Intent Tower $Z_i$:} 
We adopt a hybrid intent representation to balance identity specificity and semantic representation. For each intent, an ID Embedding $\mathbf{z}_{id}$ captures historical interaction patterns via learnable lookup tables, while a semantic embedding $\mathbf{z}_{sem}$ is extracted from textual descriptions using a pretrained HLLM~\cite{hllm} intent encoder. These embeddings are fed into a feature cross layer to produce a unified vector $\mathbf{z}_{i}$. To ensure production-level efficiency, all $\mathbf{z}_{i}$ are pre-computed and cached in an Offline Index. During inference, the system performs a constant-time embedding lookup instead of real-time encoding, significantly boosting throughput and reducing latency.

\textbf{Query Tower $Z_q$:} 
A single-layer MLP projects the semantic embedding $e_{LTP}$ from LLM to ranking-aware embedding $\mathbf{z}_{q}$. This approach offers high computational efficiency and is particularly effective for large-scale industrial datasets.


Given context embedding from three towers, we define the ranking score as a weighted combination of user–intent and query–intent similarities:
\begin{equation}
    s = \omega \cdot sim(\mathbf{z}_u, \mathbf{z}_i) + (1-\omega) \cdot sim(\mathbf{z}_i, \mathbf{z}_q)
\end{equation}
where $\omega$ controls the relative contribution of collaborative signals and semantic matching. $sim$ represent cosine similarity function. In this formulation, the interaction between $(z_i,z_q)$ primarily models semantic relevance, while $(z_u,z_i)$ captures collaborative signals derived from user features.

Following the ListNet framework\cite{Cao2007LearningTR}, we optimize the model by minimizing the weighted Kullback-Leibler (KL) divergence over samples of size $B$. To mitigate the impact of noisy samples and inherent data bias, we introduce a confidence predictor that assigns a reliability weight $w_j$ to $j$th sample, enabling end-to-end automated debiasing:
\begin{equation}
\mathcal{L}_{\text{intent}}(\mathbf{s}, \mathbf{y}) = \sum_{j=1}^B w_j \cdot \text{KL}\left(P_{\text{click}}^{(j)} \parallel P_{\text{pred}}^{(j)}\right)
\end{equation}

Both $P_{\text{pred}}$ and $P_{\text{click}}$ are derived via a Softmax operation:
\begin{equation}
P_{\text{pred}}(j) = \frac{\exp(s_j)}{\sum_{k=1}^B \exp(s_k)}, \quad P_{\text{click}}(j) = \frac{\exp(y_j)}{\sum_{k=1}^B \exp(y_k)}
\end{equation}

\section{RGAlign-Rec: Ranking Guide Alignment}
\label{sec:method}
We propose an iterative refinement mechanism where the recommendation model and LLM mutually enhance each other through iterative optimization. Unlike prior work that treats LLM integration as a one-way optimization, our framework enables continuous improvement through closed-loop semantic feedback.

\begin{figure}[h]
\centering
\includegraphics[width=0.48\textwidth]{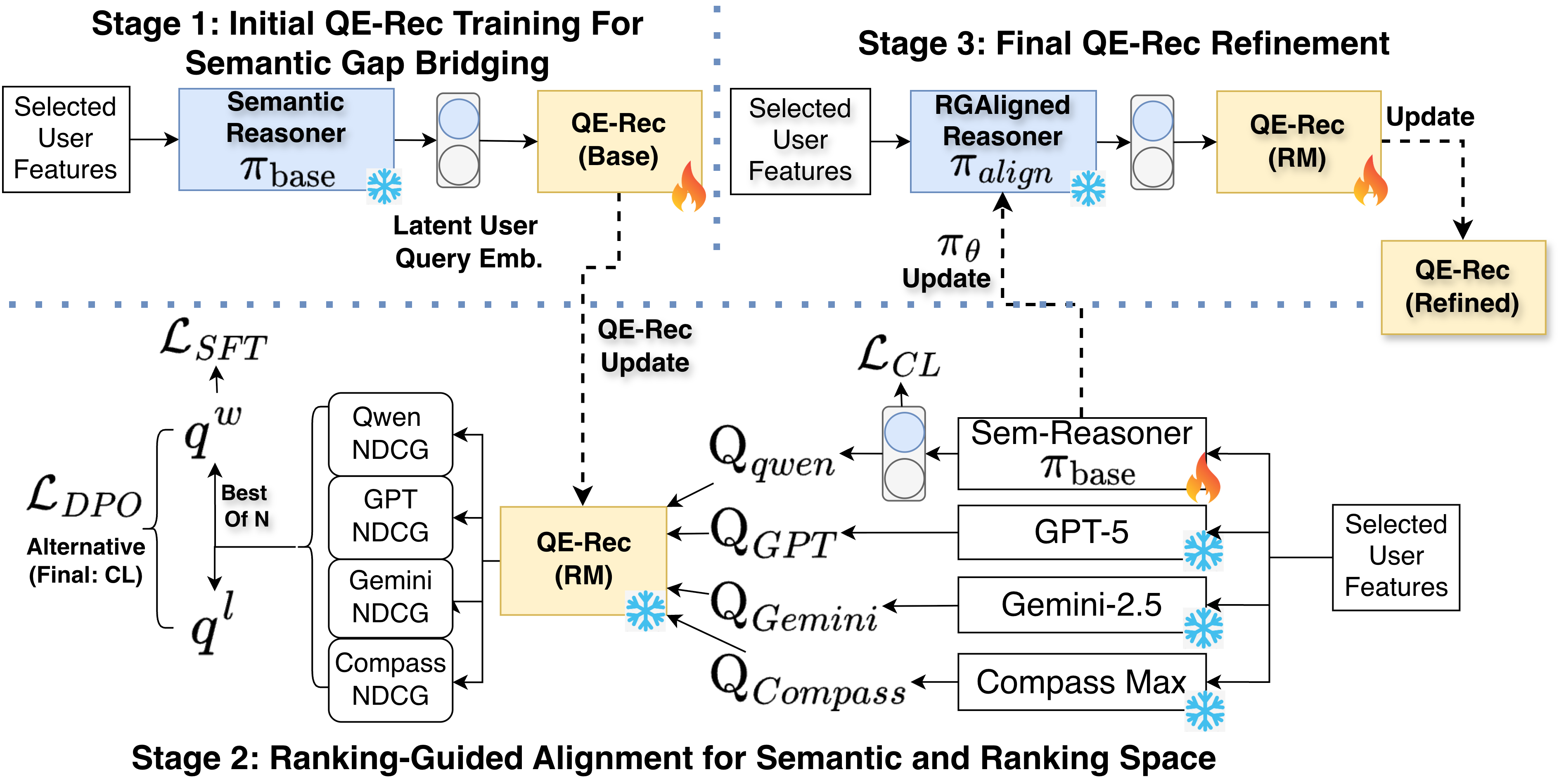}
\caption{RGAlign-Rec Training Pipeline}
\label{fig:training pipeline}
\end{figure}

The framework consists of three stages shown in Figure \ref{fig:training pipeline}:

\subsection{Stage 1: Initial QE-Rec Training.} To establish a robust evaluation mechanism for semantic alignment, we first train a QE-Rec model to serve as a stable Reward Model. This decoupled preliminary stage is essential because direct semantic alignment of the LLM is unattainable without a reliable reward signal; furthermore, simultaneous joint training with a pre-trained LLM frequently induces significant model instability. Central to this phase is the use of a frozen semantic reasoner ($\pi$), which transforms raw user features into latent semantic embeddings. These embeddings are then fed into the query tower and integrated with the intent and user tower to facilitate model training with loss function detailed in section \ref{subsec: QE-rec}. This process ensures a consistent and grounded foundation for the subsequent semantic alignment of the primary LLM.

Train the Query-Enhanced recommendation model with LTP embeddings from a frozen LLM:
\begin{equation}
    \delta^*_{\text{RCMD}} = \arg\min_\theta \mathcal{L}_{\text{rank}}(\theta; \mathbf{e}_{\text{LTP}})
\end{equation}

While the initial QE-Rec training successfully bridges the semantic gap inherent in traditional two-tower ranking models in our zero-query recommendation scenario, the semantic embeddings remain anchored in the general LLM's latent space. To ensure these representations are effective for recommendations, the subsequent stage focuses on ranking-guided alignment. This involves mapping general-purpose embeddings into the ranking-specific manifold and injecting specialised e-commerce domain knowledge.

\subsection{Stage 2: Ranking-Guided Alignment} In this stage, we discuss alignment methods to enhance the latent query reasoning capabilities of general LLM and detail how to utilize QE-Rec as a reward model to better align the ranking and semantic spaces.

\textbf{Multi-LLM Sample Generation} To bolster latent query reasoning, we adopt a multi-LLM distillation strategy to distil reasoning knowledge from the teacher model. First, leverage 3 industry-leading LLMs (Gemini-2.5-Pro/GPT-5/CompassMax~\cite{compassmax}) plus LLM reasoner (qwen3-4B) to generate users' latent queries. Given that the internal "last token" embeddings of these proprietary models are inaccessible, we employ the qwen3-4b embedding model to map these queries into a unified semantic embedding set. This ensures consistency with the QE-Rec Query Tower's input format and facilitates direct comparison within the same semantic space.

Subsequently, we pair these semantic embeddings with the original user data, effectively quadrupling the user-intent-query samples. We then employ the QE-Rec model from Stage 1 as a reward model to evaluate these query-enhanced embedding sets, generating ranking scores for each candidate produced by the 4 LLMs. 

\textbf{Best-of-N Sampling Strategy} To quantify the quality of each candidate query, we adopt NDCG~\cite{ndcg} as the primary metric, as it effectively captures both the relevance of the recommendation results and the precision of their ranking positions. We designed four progressive sampling strategies to balance query quality, model diversity, and data scale.

\begin{enumerate}
    \item \textbf{Full Retention (V1):} For each sample, the query with the highest NDCG score is selected. In the event of a tie in NDCG scores, selection follows a predefined Priority: \\ $\text{CompassMax} > \text{Gemini} > \text{GPT-5} > \text{Qwen3}$
    \item \textbf{Non-Qwen3 Filtering (V2):} Follows the same selection rules as V1 with an additional constraint: if the optimal query from the baseline Qwen3-4B model, the sample is discarded.
    \item \textbf{Intersection Filtering (V3):} We select candidates that maximize Edit Distance from the Qwen3-4B generated query (for diversity) and exhibit the highest Cosine Similarity to the golden intent (for accuracy). Only candidates satisfying both criteria are retained.
    \item \textbf{Union Filtering (V4):} Similar to Strategy 3, but employs a union operation to retain samples satisfying either criterion, thereby increasing data diversity and coverage. 
\end{enumerate}
Finally, by implementing various Best-of-N sampling strategies, we identify the optimal queries $\mathbf{Q^{w}}$ and others as negative samples $\mathbf{Q^{l}}$ to serve as the ground truth for the post-training phase.



\textbf{Ranking-Guided Alignment (RGA):} To effectively transfer the feedback signals from QE-Rec into the LLM, we design a two-stage ranking guided alignment protocol. The first stage establishes foundational query generation capabilities through \textbf{Ranking Guided SFT (RG-SFT)}, followed by a second stage of Ranking Guided Refinement to further refine generation quality for ranking model.

The goal of RG-SFT is to train the LLM to generate queries guiding the ranking model achieve better CTR prediction. Given the user features and context $C$ and the optimal query selected via QE-Rec feedback $q^{w} \in Q^{w}$ as ground truth, the RG-SFT optimization objective is defined as:
\begin{equation}
L_{RG-SFT} = - \mathbb{E}_{(C, q^{w}) \sim D} \left[ \sum_{t=1}^{|q^{w}|} \log \mathbf{\pi}_{\theta}(q^{w}_t \mid c, q^{w}_{<t}) \right]
\end{equation}

where $\mathbf{\pi}_{\theta}$ denotes the parameterized Semantic Reasoner and $\theta$ represents the its parameters. In our RGA framework, RG-SFT serves as a distillation mechanism that injects downstream ranking signals—captured by the QE-Rec reward model—directly into the LLM’s generative policy. By fine-tuning the optimal queries identified through Best-of-N ranking results, the model learns to bridge the gap between general semantic representation and ranking performance, ultimately internalizing a strategy that optimizes for both linguistic diversity and recommendation accuracy.

Building upon the model trained on RG-SFT, we employ two advanced ranking guide refinement strategies: pregerence-based (RG-DPO) and representation-level (RG-CL) alignment to optimize model performance separately. 

\textbf{RG-DPO:} Formulating a preference-based learning objective to maximize the margin between optimal $q^{w}$ and suboptimal queries $q^{l} \in Q^l$.

The preference pairs are constructed as:
\begin{equation}
\mathcal{D}_{\text{pref}} = \{(\mathcal{C}_i, q_i^{w}, q_i^{l}) \mid \text{NDCG}(q_i^{w}) > \text{NDCG}(q_i^{l})\}
\end{equation}

The optimization objective is formulated as:
\begin{equation}
\begin{aligned}
\mathcal{L}_{\text{DPO}}(\theta)
= -\mathbb{E}_{(\mathcal{C}, q^{w}, q^{l}) \sim \mathcal{D}_{\text{pref}}}
\Bigg[
\log \sigma \Bigg(
& \beta \log \frac{\pi_\theta(q^{w} \mid \mathcal{C})}{\pi_{\text{ref}}(q^{w} \mid \mathcal{C})} \\
& {}- \beta \log \frac{\pi_\theta(q^{l} \mid \mathcal{C})}{\pi_{\text{ref}}(q^{l} \mid \mathcal{C})}
\Bigg)
\Bigg]
\end{aligned}
\end{equation}
where $\pi_\theta$ is the policy model that is being optimized, $\pi_{\text{ref}}$ is the frozen version of the SFT checkpoint, which serves as a constraint to prevent the model from drifting away from the distribution of the learned instruction follow-up during the refinement phase. $\beta$ is a temperature parameter controlling the strength of the optimization and $\sigma$ is the Sigmoid function.

RG-DPO enhances the model's discriminative power by explicitly maximising the reward margin between ranking-optimal and suboptimal queries, transitioning from simple behavioural cloning to ranking-aware decision-making. By utilizing the SFT model as a reference, it stabilizes the alignment of the LLM’s output distribution with the expert preferences of QE-Rec while preserving foundational linguistic capabilities. However, its effectiveness depends entirely on the diversity of the Best-of-N candidate pool, as insufficient (only 3 in our scenario) or low-quality negatives can lead to noisy gradients or trivial learning. Furthermore, DPO focuses on optimizing token-level probabilities rather than latent geometric alignment, which may result in reward hacking where the model finds surface-level shortcuts rather than achieving deep semantic synchronization with the ranking space.

\textbf{RG-CL} addresses the latent geometric gap of DPO by directly synchronizing the LLM's internal hidden states with the ranking-optimized semantic space of the expert queries. By enforcing feature-level consistency, it ensures that the model’s last-token embeddings are inherently calibrated to the downstream ranking preferences, facilitating more robust intent representation. 

To bridge the gap between the generative hidden space and the ranking-optimized semantic space, we employ an InfoNCE-based contrastive objective. For a given batch of size $N$, let $\mathbf{h}_i$ be the last-token hidden state of the LLM for the $i$-th user sample, and let $\mathbf{e}_i^+$ be the semantic embedding of the optimal query (highest NDCG from QE-Rec) as golden label. The loss function is formulated as:
\begin{equation}
L_{CL} = - \frac{1}{N} \sum_{i=1}^{N} \log \frac{\exp(\text{cos}(\mathbf{h}_i, \mathbf{e}_i^+) / \tau)}{\sum_{j=1}^{N} \exp(\text{cos}(\mathbf{h}_i, \mathbf{e}_j^+) / \tau)}
\end{equation}
where $\tau$ is the temperature hyperparameter that scales the sharpness of the similarity distribution.

\subsection{Stage 3: Closed-Loop Calibration} 
To finalize the alignment, we implement a Closed-Loop Calibration stage that updates the QE-Rec model using the refined query distribution from the aligned LLM $\pi_{align}$. This iterative process mitigates potential semantic drift and ensures that the recommendation model is perfectly synchronized with the LLM’s enhanced reasoning capabilities. Once the mutual alignment between the generative and ranking spaces is achieved, the integrated RGAlign-Rec framework is deployed for production serving, delivering robust and ranking-optimized recommendations.



\section{Experiments}
\label{sec:experiments}
To validate the effectiveness of RGAlign-Rec, we first conduct extensive offline experiments based on 5-week user interaction logs and user clicking feedback within Shopee Chatbot, an e-commerce customer service platform serving billion-scale users across Southeast Asia and Latin America. Furthermore, we conduct extensive experiments to answer the following research questions:
\begin{itemize}
    \item \textbf{RQ1:} How does RGAlign-Rec perform compared to production baselines?
    \item \textbf{RQ2:} What is the contribution of LTP embeddings vs. traditional embeddings?
    \item \textbf{RQ3:} How does the reward sampling strategy affect model performance?
\end{itemize}
Finally, we deploy our method online and reveal its benefit in online A/B tests.

\subsection{Experimental Setup}

\textbf{Dataset.} 
The datasets were constructed from online interaction logs and user feedback across eight different regions. They consist of 89 features, covering millions of users and thousands of intents. The data span a five-week period, totalling around 7.53M samples, with the first four weeks used for training and the final week reserved for validation. Click signals are directly adopted as training labels. For the evaluation set, totalling around 46K, human annotators are involved to refine the labels into users’ true intents as click signals, aiming to mitigate the noise inherent in implicit feedback.


\textbf{Metrics.} 
We employ four popular metrics for offline evaluation: GAUC, Recall@K, NDCG@K, and MRR, for $K \in \{3, 5\}$. For online metrics, we use top-K CTR (CTR@K) and Intent Hit Rate(IHR) for model evaluation and CSAT as business metrics. Specifically, we define Intent Hit Rate (IHR) using an LLM-as-a-judge approach to measure real-world effectiveness. An offline LLM (GPT-5) analyses full session transcripts—including both chatbot and agent interactions—to identify the user’s True Intent. A "hit" is recorded if this identified intent appears within the model's Top-3 recommendations. The metric is calculated as:$$IHR = \frac{1}{N} \sum_{i=1}^{N} \mathbb{1}(\text{True Intent}_i \in \text{Top-3 Intent}_i)$$where $N$ is the total number of sessions and $\mathbb{1}(\cdot)$ is the indicator function that equals 1 if the condition is met, and 0 otherwise. CSAT is defined as a good ratings of users divided by all good and bad ratings to evaluate Chatbot servide quality.


\textbf{Base Model} Our baseline utilises a two-tower architecture integrating AutoInt~\cite{Song2018AutoIntAF} with a Multi-Gate Mixture-of-Experts (MoE) framework, representing a mature SOTA paradigm in industrial recommendation. This model explicitly captures high-order feature interactions via AutoInt’s self-attention mechanism, employing a 4-expert dense MoE structure to navigate the heterogeneous intent distributions found in intelligent customer service. By separately modeling user and intent representations before joint label prediction, this design maintains high predictive accuracy while efficiently scaling to diverse production scenarios.

\textbf{Implementation.} 
Our system is optimized for production using 8-bit quantization on a single NVIDIA B200 GPU, supporting 50 peak QPS. The LLM reasoner and overall RGAlign-Rec service exhibit latencies of 120ms and 1.5s, which are well within the acceptable response threshold for Chatbot services. For the QE-Rec component, we utilize a weighted fusion strategy that assigns a $0.7$ weight to the user-intent collaborative logits. The model is trained for $10$–$20$ epochs with a dropout rate of $0.2$, weight decay of $0.01$, and gradient norms clipped at $1.0$. The SFT-CL phase utilizes a joint multi-objective loss, combining contrastive learning ($\lambda_{\text{CL}} = 1.0$) and supervised causal language modeling ($\lambda_{\text{causal}} = 0.01$), with a temperature $\tau = 0.05$ to sharpen similarity distributions. To manage memory constraints during long-context training, we utilize gradient accumulation to achieve an effective batch size of $24$ and implement DeepSpeed ZeRO Stage-3 for efficient model state partitioning. The learning rate is set to $5 \times 10^{-5}$ with a linear decay schedule and a $3\%$ warm-up phase over $10$ epochs. Finally, we apply NEFTune-based noise regularisation to the embedding layer ($\alpha = 5$) to enhance training robustness and mitigate overfitting.

\subsection{Offline Performance Comparison (RQ1)}
To evaluate the effectiveness of the proposed RGAlign-Rec framework, 
Table~\ref{tab:main_results} presents the main experimental results on Shopee dataset comparing RGAlign-Rec model variants against the production baseline.

\begin{table*}[h]
\caption{Main results on Shopee dataset. All improvements are relative to the production baseline. Best results in \textbf{bold}. R@K=Recall@K, N@K=NDCG@K.}
\label{tab:main_results}
\vspace{-3mm}
\begin{tabular}{llllcccccc}
\toprule
Model & Stage1(QE) & Stage2(RGA) & Stage3 & GAUC & R@3 & R@5 & N@3 & N@5 & MRR \\
\midrule
Baseline & - & - & - & 96.31 & 67.73 & 76.81 & 70.84 & 74.96 & 58.52 \\
\midrule
QE-Rec & Raw LLM & - & No & 96.42 & 67.99 & 77.09 & 71.04 & 75.07 & 58.79 \\
RGAlign-Rec & Tuned & SFT\&CL & No & 96.42 & 68.07 & 77.13 & 71.07 & 75.11 & \textbf{58.84} \\
RGAlign-Rec & Tuned & SFT\&DPO & Yes &96.42 & 68.06 & 77.13 & 71.09 & 75.14 & 58.82 \\
RGAlign-Rec(Full) & Tuned & SFT\&CL & Yes & \textbf{96.43} & \textbf{68.11} & \textbf{77.21} & \textbf{71.09} & \textbf{75.14} & 58.81 \\
\midrule
\multicolumn{4}{l}{\textit{Rel. Improv. (Best)}} & +0.12\% & +0.56\% & +0.52\% & +0.35\% & +0.24\% & +0.50\% \\
\multicolumn{4}{l}{\textit{Error Reduction (Best)}} & +3.52\% & +1.18\% & +1.72\% & +0.86\% & +0.72\% & +0.70\% \\
\bottomrule
\end{tabular}
\end{table*}


Overall, RGAlign-Rec (Full) consistently outperforms the production baseline across all key metrics. Specifically, it achieves a +0.12\% absolute gain in GAUC, representing a significant 3.52\% error reduction, and substantial improvements in ranking accuracy with +0.56\% in R@3 and +0.35\% in N@3. These results demonstrate the framework's effectiveness in leveraging LLM as a semantic reasoner to enhance recommendation utility in a large-scale industrial setting. The evaluation results in various training stage reveal several interesting observations, as outlined below:
\begin{itemize}
    \item The Query-Enhanced recommendation (QE-Rec) yields an immediate performance lift both on all metrics. This validates that the rich semantic knowledge within general LLMs can provide valuable latent query signals that traditional ranking models lack, and our query-enhanced model can integrate the semantic info for ranking effectively. However, the gains from Raw LLM integration alone are limited, as the LLM's general-purpose output is not yet aligned for the specific ranking objectives of a recommendation system.
    
    \item In Stage 2, compared with the Stage 1 QE-Rec, RGAlign-Rec (SFT+CL) further increases the MRR from 58.79\% to 58.84\% and improves Recall@3 by 0.07\%, even without iterative refinement. These results demonstrate that in-batch contrastive learning effectively synchronises the LLM’s latent space with the ranking-optimised semantic space by enforcing geometric consistency.
    
    \item Finally, representation-level alignment (SFT+CL) exhibits greater robustness and scalability during iterative refinement compared to preference-based methods(SFT+DPO). While the DPO-based calibration provides competitive NDCG scores, the RGAlign-Rec (Full) model—which utilises SFT+CL followed by calibration—achieves the best overall results in GAUC (96.43) and Recall (68.11 R@3, 77.21 R@5). This indicates that representation-level alignment is more receptive to iterative updates, allowing the model to more robustly resolve objective misalignment between the reasoner and the ranker. 
\end{itemize}
Consequently, we select the RGAlign-Rec(Full) framework for final production deployment, as it demonstrates the most consistent performance gains and structural adaptability in the closed-loop refinement process.

\subsection{Ablation Study}
\subsubsection{Pooling Strategy (RQ2)}
We conducted an ablation study to evaluate the performance of the Last Token Pooling (LTP) embedding compared to BGE embedding mapping(BGE). As shown in the Table~\ref{tab:lpt_ablation}, LTP consistently outperforms BGE across all stages and metrics. At stage 3, LTP achieves a significant improvement over BGE, with relative gains of +0.81\% in Recall@3 and +0.77\% in MRR. These results demonstrate that the last token serves as a superior semantic bottleneck, effectively capturing the global context of the sequence in a decoder-only architecture. The inferior performance of BGE is attributed to its two-stage process, where the LLM first generates a query that is subsequently mapped by a frozen BGE model. This introduces a static external variable that remains unoptimized during Stage 2 alignment, creating a discrepancy that degrades objective alignment. Conversely, LPT directly leverages the LLM’s internal hidden states, facilitating seamless end-to-end optimisation of the ranking objective. This allows the LLM reasoner to focus on ranking-aware semantic signals within the latent space, without the representation bottleneck of a frozen encoder.

\begin{table}[h]
\caption{Ablation study on embedding methods. LTP significantly outperforms traditional embeddings.}
\label{tab:lpt_ablation}
\vspace{-3mm}
\resizebox{0.48\textwidth}{!}{\begin{tabular}{llccccccc}
\toprule
 Model & Pooling Strategy & GAUC & R@3 & R@5 & N@3 & N@5 & MRR \\
\midrule
QE-Rec & BGE & 96.39 & 67.93 & 77.04 & 70.94 & 75.01 & 58.67 \\
QE-Rec & LTP & 96.42 & 67.99 & 77.09 & 71.04 & 75.07 & 58.79 \\
\midrule
RGAlign-Rec(Full) & BGE & 96.31 & 67.56 & 76.69 & 70.69 & 74.80 & 58.36 \\
RGAlign-Rec(Full) & LTP & \textbf{96.43} & \textbf{68.11} & \textbf{77.21} & \textbf{71.09} & \textbf{75.14} & \textbf{58.81} \\
\midrule
\multicolumn{2}{l}{\textit{LTP vs BGE (Rel.)}} & +0.12\% & \textbf{+0.81\%} & +0.68\% & +0.57\% & +0.45\% & +0.77\% \\
\bottomrule
\end{tabular}}
\end{table}





\subsubsection{Best-of-N Sampling Strategy (RQ3)}
The sampling strategy serves as our unified mechanism for semantic representation across all training stages. As shown in Table \ref{tab:sampling}, Strategy V1 (Full Retention) performs optimally when the backbone is frozen (QE-Rec). In this static setting, including all candidates—including those from the LLM reasoner (Qwen3-4B)—provides the most comprehensive feature set for the ranker to evaluate.

However, during stages 2-3, when the LLM reasoner is updated, V1’s performance saturates early. This is primarily due to self-fitting bias: by training on its own generated queries, the model tends to over-optimise toward its internal distribution rather than the ground truth ranking objective. To mitigate this, Strategy V2 (Non-Qwen3) explicitly excludes queries from the LLM reasoner, forcing the system to align with diverse, high-quality ranking-aware signals from external teacher LLMs to improve the reasoner's performance. This approach acts as a robust form of cross-model regularization, preventing the feedback loop inherent in V1. Consequently, V2 facilitates superior objective alignment, achieving the highest Recall@3 (68.11\%) and Recall@5 (77.21\%) in the final iterative refinement stage. Based on these findings, we adopt V2 to ensure long-term scalability and generalizability.

\begin{table}[h]
\caption{Ablation on Best-of-N sampling strategies at Stage 2.}
\label{tab:sampling}
\vspace{-3mm}
\resizebox{0.48\textwidth}{!}{\begin{tabular}{llccccccc}
\toprule
Model & Strategy & GAUC & R@3 & R@5 & N@3 & N@5 & MRR \\
\midrule
QE-Rec & V1 (Full Retention) & \textbf{96.44} & 68.08 & 77.19 & \textbf{71.11} & \textbf{75.17} & \textbf{58.84} \\
QE-Rec & V2 (Non-Qwen3) & 96.42 & 68.07 & 77.13 & 71.07 & 75.11 & \textbf{58.84} \\
QE-Rec & V3 (Intersection) & 96.41 & 68.05 & 77.13 & 71.04 & 75.08 & 58.79 \\
QE-Rec & V4 (Union) & 96.42 & 68.05 & 77.18 & 71.08 & 75.12 & 58.81 \\
RGAlign-Rec(Full) & V1 & 96.42 &	68.05 &	77.16 &	71.06 &	75.11 &	58.83 \\
RGAlign-Rec(Full) & V2 & 96.43 &	\textbf{68.11} &\textbf{77.21} &	71.09 &	75.14 &	58.81 \\
\bottomrule
\end{tabular}}
\end{table}



\subsection{Online Deployment and A/B Test}
To meet the stringent throughput and latency requirements of an industrial production environment, we deploy our model on a B200 GPU using the vLLM serving framework, together with 8-bit quantisation to reduce memory footprint and inference latency while maintaining ranking accuracy.

The online A/B test results in Table~\ref{tab:abtest} demonstrate the cumulative benefits of the proposed multi-stage framework under real-world traffic. QE-Rec achieves significant gains over the production baseline, most notably a +0.98\% increase in CTR@3, illustrating its effectiveness in mitigating the semantic gap between latent user needs and intent representations in Zero-Query settings. Building upon this foundation, RGAlign-Rec(Full) further contributes an additional +0.13\% gain in CTR@3 compared to QE-Rec, indicating that ranking-guided semantic alignment enhances the model’s ability to surface the correct intent at top positions. We observe a marginal -0.21\% decline in CSAT after alignment. As CSAT reflects downstream resolution quality beyond the intent ranking stage, this divergence suggests that while the model becomes more accurate at identifying the user’s true intent, the quality of response content associated with clicked intents may remain a limiting factor that is orthogonal to ranking. Overall, the monotonic gains in ranking-oriented metrics across stages confirm the framework’s ability to capture real user intent under production constraints.


\begin{table}[h]
\caption{Online AB Test Performance for QE-Rec and RGAlign-Rec}
\label{tab:abtest}
\vspace{-3mm}
\resizebox{0.48\textwidth}{!}{\begin{tabular}{lccc}
\toprule
ABTest Setting  & CSAT & CTR@3 & IHR  \\
\midrule
QE-Rec vs Baseline & +0.17\% & +0.98\% & +1.43\% \\
RGAlign-Rec(Full) vs QE-Rec & -0.21\% & +0.13\% & +0.30\% \\
\bottomrule
\end{tabular}}
\end{table}

\section{Conclusion}
\label{sec:conclusion}



This study addresses the semantic gap in modern recommendation systems by integrating an LLM-based semantic reasoner with a query-enhanced (QE) ranking model for proactive intent prediction. To resolve the misalignment between general-purpose LLM generation and task-specific ranking objectives, we introduce the Ranking-Guided Alignment (RGA) framework, which leverages downstream ranking signals as feedback to guide semantic alignment. Through representation-level alignment and iterative refinement, RGA effectively synchronizes the latent semantic space of the LLM with the operational requirements of the ranker. Extensive offline evaluation and online deployment demonstrate that the resulting RGAlign-Rec system delivers consistent improvements in both CTR and Intent Hit Rate (IHR) in a large-scale production environment. Overall, this work presents a practical and scalable pipeline for distilling ranking-aware intelligence into generative models, enabling reliable semantic reasoning under real-world industrial constraints.

\definecolor{llmcolor}{RGB}{255, 242, 204} 
\definecolor{qecolor}{RGB}{218, 232, 252}  
\definecolor{frozen}{RGB}{245, 245, 245}   
\definecolor{shopee_orange}{RGB}{238, 77, 45} 

\bibliographystyle{ACM-Reference-Format}
\bibliography{references}

\end{document}